\documentclass[11pt, a4paper, oneside, english]{article}
\usepackage[english]{babel}
\usepackage[utf8]{inputenc}
\usepackage{hyperref}
\usepackage{authblk}
\usepackage{listings}

\usepackage{amsmath,amssymb,graphicx,bbm}

\usepackage{stix}

\usepackage[body={15.0cm, 23.0cm},top=2.0cm,bottom=2.5cm,left=2.0cm,right=2.0cm]{geometry}
\graphicspath{{images/}}

\newcommand{\assign}{:=}
\newcommand{\tmop}[1]{\ensuremath{\operatorname{#1}}}

\begin{document}

\title{Deviations from Gaussianity in deterministic discrete time dynamical
systems}

\author{
  Jeroen Wouters
}

\affil{Department of Mathematics and Statistics\\ University of
	Reading\\\texttt{j.wouters@reading.ac.uk}}

\maketitle

\begin{abstract}
  In this paper we examine the deviations from Gaussianity for two types of random variable
  converging to a normal distribution, namely sums of random variables generated by a deterministic discrete time map and a linearly damped  variable driven by a deterministic map. We demonstrate how Edgeworth expansions provide a universal description of the deviations from the limiting normal distribution. We derive explicit expressions for these asymptotic expansions and provide numerical evidence of their accuracy.
\end{abstract}

\section{Introduction}

Randomness provides a powerful way of describing the large-scale behaviour
of many systems in the natural and man-made world. Well-known examples are Brownian
particle motion, price evolution on financial markets and the evolution of the
Earth's climate. However, many of these systems are described by
deterministic dynamical systems on small scales. A natural question is then
how randomness arises from deterministic dynamics.

One much-explored way in which random variable can arise from deterministic
dynamical systems is through variations of the central limit theorem. In such
theorems, many nearly independent contributions are added or integrated over to
result in a Gaussian random variable. This principle has for example been
explored for systems with a bath of a large number of deterministic
oscillators {\cite{FordEtAl65}}. Another way to obtain sums of nearly
independent contributions is to sum over time series of sufficiently mixing
dynamical systems. The evolution is completely deterministic, with the only randomness appearing through the initial conditions. This approach has been investigated for discrete time
dynamical systems in
{\cite{denker1989central,luzzatto_stochastic-like_2006,young_recurrence_1999}}.
An extension of this case can be found in the study of slow-fast dynamical
systems where instead of simply summing the output of a dynamical system the
slow variable has a non-trivial dynamics of its own. This setting has been
studied in {\cite{MelbourneStuart11,GottwaldMelbourne13c,KellyMelbourne17}}.

In these theorems we have to consider specific limits, for example, taking the number of
oscillators, the length of \ sums or the time scale separation to inifity.
Such conditions are of course never fulfilled in reality. Therefore, the
distributions observed in a physical system will deviate from the limiting
distribution predicted by theory. These deviations can in many cases be
successfully described by Edgeworth expansions, which provide correction terms
to the limiting distribution
{\cite{fernando_edgeworth_2018,HervePene10,williams_stochastic_2018,wouters_edgeworth_2019}}. Edgeworth expansions have furthermore been used to develop reduced order models for slow-fast dynamical systems \cite{wouters_stochastic_2019}.

Here we consider two applications of Edgeworth expansions. First of all, we
describe a method to derive the Edgeworth coefficients of sums of dependent
random variables. We corroborate our results by numerical experiments. Secondly, we show
that some recent results on approximations of invariant distributions of
slow-fast discrete maps are in fact a specific case of the Edgeworth
expansion.

The article is structured as follows. In Section \ref{sec.clt} we give a brief overview of central limit theorems and the Edgeworth expansion. In Section \ref{sec.sums} we examine sums of time series of a deterministic dynamical
system with random initial conditions. In Section \ref{sec.multiscale} we
study a type of slow-fast dynamical system with linear damping of the
slow variable. We show that the deviations of the invariant measure of the
slow variable can effectively be described by an Edgeworth expansion.

\section{Central limit theorems and Edgeworth expansions}
\label{sec.clt}

The convergence of appropriately normalized sums of random variables to
Gaussian, Poisson or other infinitely divisible distributions is an important
topic in probability theory and dynamical systems theory. Theorems showing
such convergence are known as central limit theorems (CLT). A CLT holds for a
sequence of random variables $z_i$ with $i \in \mathbbm{N}$ and $\mathbbm{E}
[z_i] = 0$ if $x_n \assign \frac{1}{\sqrt{n}} \sum_{i = 1}^{n} z_i$
converges in distribution to a normal distribution $\mathcal{N}_{0, \sigma^2}$
with mean $0$ and variance $\sigma^2$ as $n \rightarrow \infty$.

If $x_n$ converges to a normal distribution $\mathcal{N}_{0, \sigma^2}$
the variance $\sigma^2$ is given by $\sigma^2 = \lim_{n \rightarrow \infty}
\mathbbm{V} [x_n]$, with $\mathbbm{V} [x_n]$ the variance of $x_n$. For a stationary
generating process $z_i$, we have
\begin{eqnarray*}
  \mathbbm{V} [x_n] & \assign & \mathbbm{E} [x_n^2] = \frac{1}{n} \sum_{i =
  0}^{n - 1} \sum_{j = 0}^{n - 1} \mathbbm{E} [z_i z_j]\\
  & = & \mathbbm{E} [z^2_1] + 2 \frac{n - 1}{n} \mathbbm{E} [z_1 z_2] + 2
  \frac{n - 2}{n} \mathbbm{E} [z_1 z_3] + \ldots + 2 \frac{1}{n} \mathbbm{E}
  [z_1 z_n],
\end{eqnarray*}
where the last equality holds by stationarity of the sequence $z_i$. Therefore
$\sigma^2$ is determined by the correlation structure of $z_i$ as
\begin{eqnarray}
  \sigma^2 & = & \mathbbm{E} [z^2_1] + 2 \sum_{i = 2}^{\infty}
  \mathbbm{E} [z_1 z_i] .  \label{eq.GK}
\end{eqnarray}
This expression is sometimes referred to as the Green-Kubo formula.

Central limit theorems have been shown to hold for i.i.d random variables
{\cite{FellerBook}}, independent but non-identical random variables
{\cite{CinlarBook}}, weakly dependent random variables {\cite{Ibragimov62}}
and deterministic discrete time maps
{\cite{denker1989central,luzzatto_stochastic-like_2006,young_recurrence_1999,bahsoun_mixing_2016,nicol_central_2018}}.
In the case of deterministic maps randomness is introduced by a random choice
of the initial condition.

Formally, the CLT can be derived by considering the characteristic function
$\chi_n (\theta) \assign \mathbbm{E} [e^{i \theta x_n}]$. By Taylor expanding
$\ln \chi_n$ in $\theta$ we have that
\begin{eqnarray*}
  \ln \chi_n (\theta) & = & \sum_{k = 0}^{\infty} \frac{c_n^{(k)}}{k!} (i
  \theta)^k
\end{eqnarray*}
where $c_n^{(k)}$ is the $k$-th cumulant of $X_n$, satisfying the recursive
relation
\begin{eqnarray*}
  c^{(k)}_n & = & m^{(k)}_n - \sum_{l = 1}^{k - 1} \binom{k - 1}{l - 1} m^{(k
  - l)}_n c^{(l)}_n \,,
\end{eqnarray*}
where $m_n^{(k)}$ is the $k$-th central moment of $x_n$: $m_n^{(k)} := \mathbbm{E} \left[ x_n^k \right]$
If it can be demonstrated that $c_n^{(2)} \rightarrow \sigma^2$ and $c_n^{(k)}
\rightarrow 0$ for $k \geqslant 3$ then $\chi_n \rightarrow e^{-
\frac{\sigma^2}{2} \theta^2}$, the characteristic function of the normal
distribution $\mathcal{N}_{0, \sigma^2}$. The convergence in distribution of
$x_n$ then follows from Lévy's continuity theorem \cite{FellerBook}.

\subsection{Deviations from the limiting distribution}

The formal derivation of the CLT in the previous section can be extended to
provide more details on the way in which the limiting distribution is approached.
This results in a so-called Edgeworth expansion, describing the deviations
from the limiting distribution in orders of $\frac{1}{\sqrt{n}}$.

We assume that the cumulants $c^{(k)}_n$ can be expanded in orders of $\sqrt{n}$ as
\begin{eqnarray}
  c^{(2)}_n & = & \sigma^2 + \frac{1}{n} c^{(2, 1)} + o \left(
  \frac{1}{n} \right) \label{eq.cum-exp} \\
  c_n^{(3)} & = & \frac{1}{\sqrt{n}} c^{(3, 1)} + o \left( \frac{1}{n} \right)
  \nonumber\\
  c_n^{(4)} & = & \frac{1}{n} c^{(4, 1)} + o \left( \frac{1}{n} \right)
  \nonumber
\end{eqnarray}
with $c^{(3, 1)}$ and $c^{(4, 1)}$ constants, and we assume that $c_n^{(p)} = o
(n)$ for $p > 4$. This assumption can be easily verified for i.i.d. random
variables and has been also shown to hold for weakly dependent random variables {\cite{GoetzeHipp83}}. These assumptions allow to expand the characteristic function

\begin{align*}
  \chi_n (\theta) & = \exp \left( c^{(2)}_n  \frac{(i \theta)^2}{2!} +
  c^{(3)}_n  \frac{(i \theta)^3}{3!} + c^{(4)}_n  \frac{(i \theta)^4}{4!} +
  \ldots \right)\\
  & = \exp \left( \frac{1}{n} c^{(2, 1)} \frac{(i \theta)^2}{2!} +
  \frac{1}{\sqrt{n}} c^{(3, 1)}_{}  \frac{(i \theta)^3}{3!} + \frac{1}{n}
  c^{(4, 1)} \frac{(i \theta)^4}{4!} + o \left( \frac{1}{n} \right) \right)
  \exp \left( - \sigma^2  \frac{\theta^2}{2} \right)\\
  & = \left( 1 + \frac{1}{\sqrt{n}} c^{(3)}_{\infty}  \frac{(i \theta)^3}{3!}
  + \frac{1}{n} \left( c^{(2, 1)} \frac{(i
  \theta)^2}{2!} + c^{(4, 1)} \frac{(i \theta)^4}{4!} + \frac{1}{2} \left(
  c^{(3, 1)} \frac{(i \theta)^3}{3!} \right)^2 \right) + o \left( \frac{1}{n}
  \right) \right)  \\
  & \hspace*{.6\textwidth} \times \exp \left( - c^{(2)}_{\infty}  \frac{\theta^2}{2}
  \right) .
\end{align*}

Since $\chi_n$ is essentially the Fourier transform of $\rho_n$, the distribution of $x_n$, an expansion in orders of $\frac{1}{\sqrt{n}}$ of $\rho_n$ can be obtained by taking the inverse Fourier transform of $\chi_n$. This results in the so-called Edgeworth expansion $\rho_n(x) = \rho_n^{(2)}(x) + o(\frac{1}{n})$ uniformly in $x$ {\cite{FellerBook}}, with
\begin{eqnarray}
  \rho^{(2)}_n (x) & = & \mathbf{n}_{0, \sigma^2} (x)  \left( 1 + \frac{c^{(3,
  1)}}{6 \sigma^3  \sqrt{n}} H_3 \left( \frac{x}{\sigma} \right) \right. \nonumber \\
  &&\left. \hspace*{\fill} + \frac{c^{(2, 1)}}{2 \sigma^2 n} H_2 \left(
  \frac{x}{\sigma} \right) + \frac{c^{(4, 1)}}{24 \sigma^2 n} H_4 \left(
  \frac{x}{\sigma} \right) + \frac{(c^{(3, 1)})^2}{72 \sigma^4 n} H_6 \left(
  \frac{x}{\sigma} \right) \right),  \label{eq.edgeworthDens}
\end{eqnarray}
where $\mathbf{n}_{0, \sigma^2} (x) = \frac{1}{\sqrt{2 \pi \sigma^2}} e^{-
\frac{x^2}{2 \sigma^2}}$ is the limiting normal distribution and $H_k (x) = (- 1)^n e^{x^2 / 2} \frac{d^n}{d x^n}
e^{- x^2 / 2}$ are Hermite polynomials. We have $\rho_n (x) = \rho^{(2)}_n (x)
+ o \left( \frac{1}{n} \right)$. This
expansion can be continued to higher orders of $\frac{1}{\sqrt{n}}$, resulting
in higher order Hermite polynomials. Note that as $n \rightarrow \infty$, we obtain the CLT result again.

\section{Sums of dynamical systems}\label{sec.sums}

In this section we examine the convergence of normalized sums $x_n \assign
\frac{1}{\sqrt{n}} \sum_{i = 1}^{n} f_1(y_i)$, where the $y_i$ are generated by
a dynamical system $y_{i+1} = g(y_i)$, with $y_1 \sim \rho_{\infty}$, i.e. the initial conditions are distributed according to the natural invariant measure of this dynamical system $\rho_{\infty}$.
In Appendix \ref{app.sums} we formally show that sums $x_n = \frac{1}{\sqrt{n}} \sum_{i=1}^n f_1 (y_i)$
 indeed follow a cumulant expansion as in Eq.~(\ref{eq.cum-exp}). We
derive explicit expressions relating the coefficients $c^{(2, 1)}$, $c^{(3,
1)}$ and $c^{(4, 1)}$ to the correlation functions of the dynamical system
$g$, supplementing the Green-Kubo formula of Eq. (\ref{eq.GK}).

We obtain
\begin{eqnarray}
  c^{(2, 1)} & = & - 2 \sum_{k = 1}^{\infty} k\mathbbm{E} [f_1 (y) f_1 (g^k
  (y))] \label{eq.cum2}\\
  c^{(3, 1)} & = & \mathbbm{E} [f_1 (y)^3] + \sum_{k = 1}^{\infty} 3
  (\mathbbm{E} [f_1 (y)^2 f_1 (g^k (y))] +\mathbbm{E} [f_1 (y)^{} f_1 (g^k
  (y))^2]) \label{eq.cum3} \\
   &&+ 6 \sum_{k = 1}^{\infty} \sum_{l = 1}^{\infty} \mathbbm{E}
  [f_1 (y) f_1 (g^k (y)) f_1 (g^{k + l} (y))] \nonumber
\end{eqnarray}
\begin{eqnarray}
  c^{(4, 1)} & = & \mathbbm{E} [f_1^4] + 4 \sum_{k = 1}^{\infty}
  \left(\mathbbm{E} \left[f_1 (y) f_1 (g^k (y))^3\right] +\mathbbm{E} \left[f_1 (y)^3 f_1 (g^k
  (y))\right]\right) \label{eq.cum4} \\
  &&+ 6  \sum_{k = 1}^{\infty} \left( \mathbbm{E} \left[f_1 (y)^2 f_1 (g^k (y))^2\right]
  -\mathbbm{E} \left[f_1 (y)^2\right]^2 \right) \nonumber\\
  &&+ 12 \sum_{k = 1}^{\infty} \sum_{l = 1}^{\infty} \left( \mathbbm{E} \left[
  f_1 (y) f_1 (g^k (y)) f_1 \left( g^{k + l} (y) \right)^2 \right]
  - \mathbbm{E} \left[f_1 (y) f_1 (g^k (y))\right] \mathbbm{E} \left[f_1^2\right] \right) \nonumber \\
  &&+ 12 \sum_{k = 1}^{\infty} \sum_{l = 1}^{\infty} \mathbbm{E} \left[f_1 (y) f_1
  (g^k (y))^2 f_1 (g^{k + l} (y))\right] \nonumber \\
  &&+ 12 \sum_{k = 1}^{\infty} \sum_{l = 1}^{\infty} \left( \mathbbm{E} \left[
  f_1 (y)^2 f_1 (g^k (y)) f_1 (g^{k + l} (y)) \right] -\mathbbm{E} \left[f_1 (y)^2\right]
  \mathbbm{E} \left[f_1 (y) f_1 (g^l (y))\right] \right) \nonumber \\
  &&+ 24 \sum_{k = 1}^{\infty} \sum_{l = 1}^{\infty} \sum_{m = 1}^{\infty}
  \left(\mathbbm{E} \left[f_1 (y) f_1 (g^k (y)) f_1 (g^{k + l} (y)) f_1 (g^{k + l + m}
  (y))\right] - \mathbbm{E}\left[f_1(y) f_1(g^k(y))\right]\mathbbm{E}\left[f_1(y) f_1(g^m(y)) \right] \right) \nonumber \\ && - 3 \sigma^4 + 6 \sigma^2 c^{(2, 1)} \nonumber
\end{eqnarray}
Here the expectation value $\mathbbm{E}$ is taken with respect to the physical invariant measure $\rho_{\infty}$ of $y_{i+1} = g(y_i)$. Note that these correction coefficients involve higher-order correlation
functions when compared to the Green-Kubo equation (\ref{eq.GK}).

We remark that our equation \eqref{eq.cum4} for $c^{(4, 1)}$ differs substantially from
the one given in {\cite{GoetzeHipp83}} without derivation. The numerical
experiments described in Section \ref{sec.sums-num} show an excellent
agreement with our equations, but not with those in {\cite{GoetzeHipp83}}.

\subsection{Numerical experiment}\label{sec.sums-num}

We verify the expansion (\ref{eq.edgeworthDens}) for the case where $y_i$ is
generated by the deterministic tripling map $y_{i + 1} = 3 y_i \tmop{mod} 1$
and $f_1 (y) = y^5 + y^4 - \frac{1}{6} - \frac{1}{5}$. The invariant measure
$\rho_{\infty}$ of the tripling map is the uniform distribution on $[0, 1]$,
therefore $\mathbbm{E} [f_1] = 0$.

The expansion coefficients $c^{(2,1)}$, $c^{(3,1)}$ and $c^{(4,1)}$ can be explicitly computed in this case by
iterating over the Markov partitions of the tripling map. A code listing to perform this calculation in the open-source mathematics software system SageMath \cite{sagemath} can be found in Appendix \ref{app.sage}.

Figure \ref{fig.tripling_hist} demonstrates the approximation of histograms of
sums $x_n=\frac{1}{\sqrt{n}} \sum_{i=1}^{n} f_1(y_i)$ of the tripling map with the approximation by both the CLT and the
Edgeworth expansion. The Edgeworth expansion clearly approximates the true
histogram much closer.

\begin{figure}[h]
	\centering
  \resizebox{175pt}{!}{\includegraphics{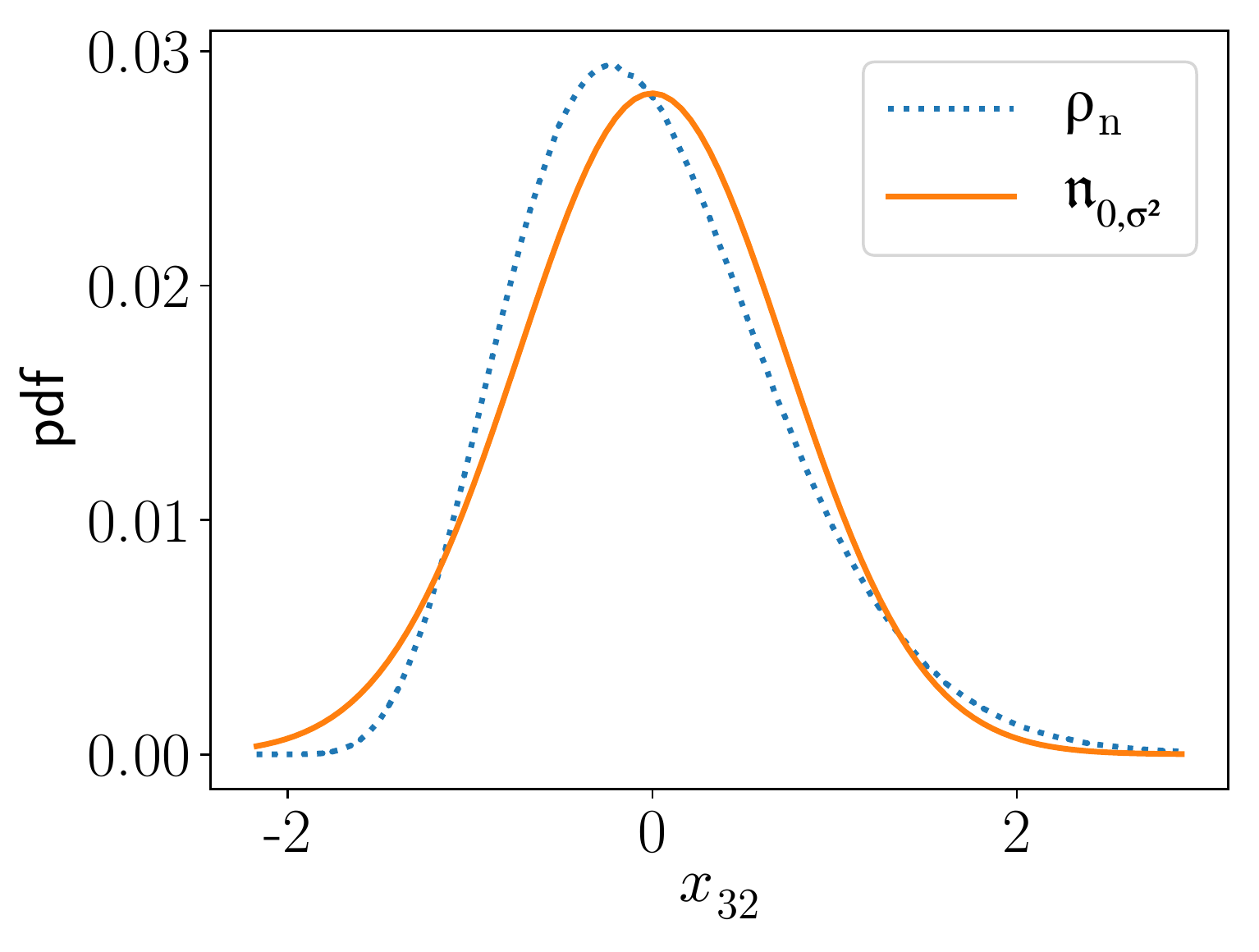}}\resizebox{175pt}{!}{\includegraphics{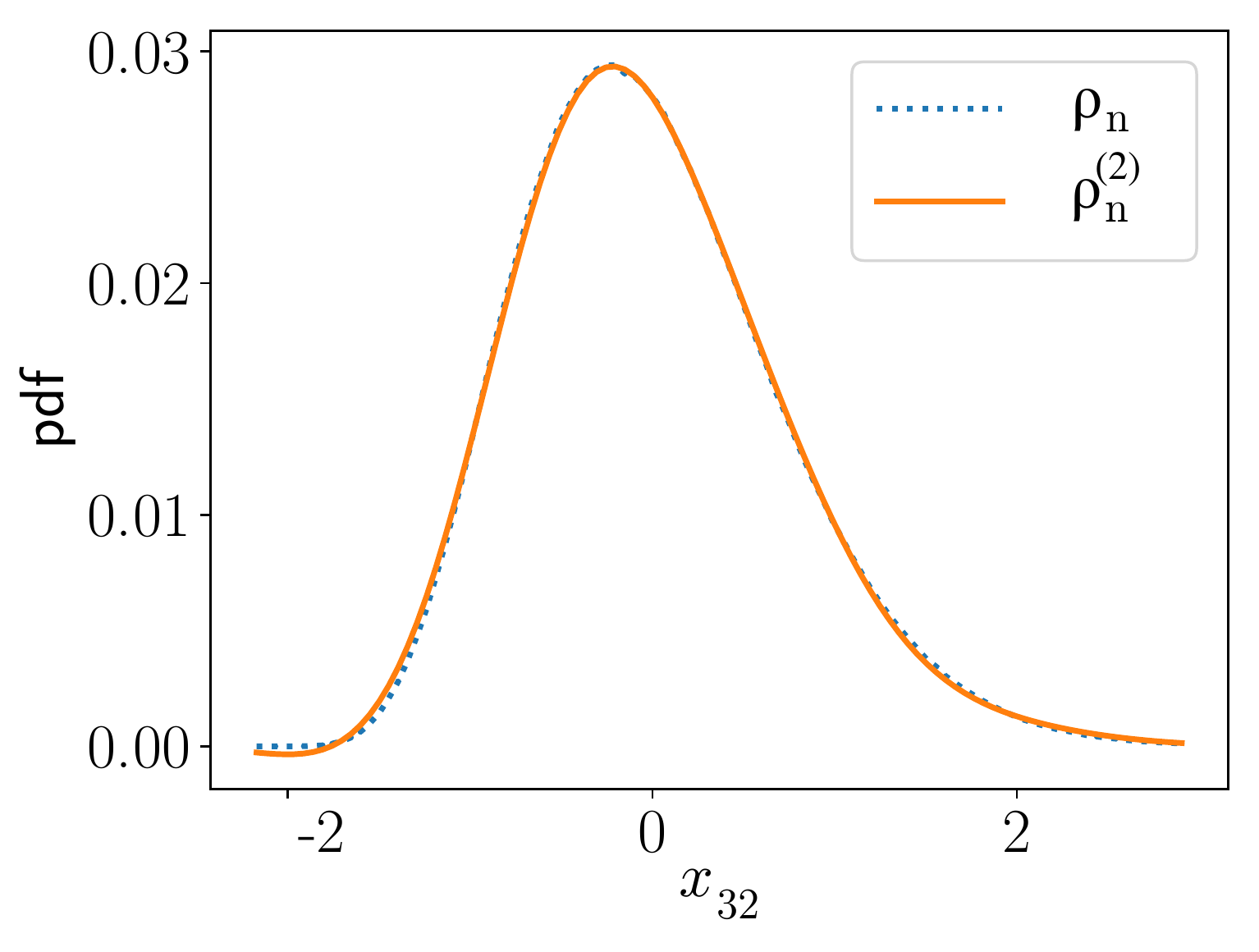}}
  \caption{The probability density of sums $\frac{1}{\sqrt{n}} \sum^n_{i = 1} f_1 (y_i)$
  with $n = 32$, $f_1 (y) = y^5 + y^4 - \frac{1}{6} - \frac{1}{5}$ and $y_i$
  generated by the tripling map from initial conditions $y_1$ uniformly
  distributed on $[0, 1]$ ($\rho_n$, dashed line) using $10^7$ samples.
  Compared to the Gaussian distribution of the Central Limit Theorem
  ($\mathfrak{n}_{0, \sigma^2}$, left figure, solid line) and the second order
  Edgeworth expansion ($\rho_n^{(2)}$, right figure, solid
  line).\label{fig.tripling_hist}}
\end{figure}

\section{Linearly damped multi-scale systems}\label{sec.multiscale}

We now consider dynamical systems of the linear Langevin type, where the deterministic output $y_n$ is not simply summed, but an additional damping term is introduced as
\begin{eqnarray}
  x_{n + 1} & = & \lambda x_n + \sqrt{\tau} y_n  \label{eq.lang2}\\
  y_{n + 1} & = & g (y_n)  \label{eq.lang1}
\end{eqnarray}
where $\lambda = e^{-\tau}$, the expectation value of $y_n$ w.r.t. the invariant
measure $\rho_{\infty}$ of $g$ is zero and we will take the limit $\tau
\rightarrow 0$. These maps have been studied in
{\cite{williams_stochastic_2018,beck_dynamical_1996}} and are a specific case
of the slow-fast maps considered in {\cite{GottwaldMelbourne13c}}\footnote{In the notation
of {\cite{GottwaldMelbourne13c}}, $\tau = \varepsilon^2$, $f_0 (y) = y$ and $f
(x, y, \varepsilon) = \frac{e^{- \varepsilon^2} - 1}{\varepsilon^2} x$.}. As
demonstrated in {\cite{GottwaldMelbourne13c}}, as $\tau \rightarrow 0$, the
paths of $y$ converge weakly to an Ornstein-Uhlenbeck process ${\mathrm d} X
= - X {\mathrm d} t + \sigma {\mathrm d} W$, where $\sigma^2$ is the
Green-Kubo variance $\sigma^2 =\mathbbm{E} [y^2_0] + 2 \sum_{i = 1}^{\infty}
\mathbbm{E} [y_0 y_i]$. Specifically, the invariant measure of $x$ converges
to a Gaussian distribution. We will now study the deviations of this measure
from the Gaussian distribution for small but non-zero $\tau$.

\subsection{Limiting distribution}

For the system (\ref{eq.lang2})-(\ref{eq.lang1}), the dependence of $x_n$ on
the history of the deterministic noise $y_i$ can be made explicit by iterating
Eq. (\ref{eq.lang2}). We get $x_n = \lambda^n x_0 + \sum_{i = 0}^{n - 1}
\sqrt{\tau} \lambda^i y_{n - 1 - i}$. In the limit $n \rightarrow \infty$ the
impact of the initial condition $x_0$ will disappear exponentially fast as
$\lambda^n$. By a change of time $i \rightarrow i - n + 1$ we are left to
consider the distribution of $x_{\infty} = \sum_{i = 0}^{\infty} \sqrt{\tau}
\lambda^i y_{- i}$.

An expression for the variance of the limiting invariant measure is
easily obtained, since
\begin{eqnarray*}
  \mathbbm{E} [x^2_{\infty}] & = & \sum_{i, j = 0}^{\infty} \tau \lambda^{i +
  j} \mathbbm{E} [y_{- i} y_{- j}]\\
  & = & \sum_{i = 0}^{\infty} \tau \lambda^{2 i} \mathbbm{E} [y_{- i} y_{-
  i}] + \sum_{k = 1}^{\infty} \sum_{i = 0}^{\infty} \tau \lambda^{2 i + k}
  \mathbbm{E} [y_{- i} y_{- i - k}] + \sum_{k = 1}^{\infty} \sum_{i =
  0}^{\infty} \tau \lambda^{2 i + k} \mathbbm{E} [y_{- i - k} y_{- i}]\\
  & = & \tau \frac{1}{1 - \lambda^2} \mathbbm{E} [y^2_0] + 2 \tau \sum_{k =
  1}^{\infty} \frac{\lambda^k}{1 - \lambda^2} \mathbbm{E} [y_0 y_k] .
\end{eqnarray*}
Taking the limit $\tau \rightarrow 0$, we obtain
\begin{eqnarray*}
  \sigma^2_{\infty} & = & \lim_{\tau \rightarrow 0} \mathbbm{E} [x^2_{\infty}]
  = \frac{1}{2} \mathbbm{E} [y^2_0] + \sum_{k = 1}^{\infty} \mathbbm{E} [y_0
  y_k] .
\end{eqnarray*}

\subsection{Corrections to the limiting distribution}

A similar calculation allows us to obtain the first Edgeworth correction term.
Calculating the third cumulant of the invariant distribution, we get
\begin{eqnarray*}
  \mathbbm{E} [x^3_{\infty}] & = & \sqrt{\tau}^3 \sum_{i, j, k = 0}^{\infty}
  \lambda^{i + j + k} \mathbbm{E} [y_{- i} y_{- j} y_{- k}]\\
  & = & \sqrt{\tau}^3 \mathbbm{E} [y^3_0] \frac{1}{1 - \lambda^3} + 3
  \sqrt{\tau}^3 \sum_{j = 1}^{\infty} \mathbbm{E} [y^2_0 y_j] \frac{\lambda^{2
  j}}{1 - \lambda^3} \\
  &&+ 3 \sqrt{\tau}^3 \sum_{i = 1}^{\infty} \mathbbm{E} [y_0 y^2_i]
  \frac{\lambda^i}{1 - \lambda^3} + 6 \sqrt{\tau}^3 \sum_{i, j = 1}^{\infty}
  \mathbbm{E} [y_0 y_i y_j] \frac{\lambda^{2 j + i}}{1 - \lambda^3}
\end{eqnarray*}
and in the limit $\tau \rightarrow 0$
\begin{eqnarray}
  c^{3,1} = \lim_{\tau \rightarrow 0} \frac{1}{\sqrt{\tau}} \mathbbm{E} [x^3_{\infty}] &
  = & \frac{1}{3} \mathbbm{E} [y^3_0] + \sum_{j = 1}^{\infty} \mathbbm{E}
  [y^2_0 y_j] + \sum_{i = 1}^{\infty} \mathbbm{E} [y_0 y^2_i] + 2 \sum_{i, j =
  1}^{\infty} \mathbbm{E} [y_0 y_i y_j] \label{eq.c31_langevin}.
\end{eqnarray}

\subsection{Numerical experiments}

Here we consider the second order Chebyshev map $g (y) = 2 y^2 - 1$. For
this map, we have that $\mathbbm{E} [y_0 y_i] = \frac{1}{2} \delta_{0, i}$, so
$\sigma^2_{\infty} = \frac{1}{4}$. The map is conjugate to the Bernoulli shift
by $y_0 = \cos (\pi u) = (\exp (i \pi u) - \exp (- i \pi u)) / 2$. Iterates
are given by $y_n = \cos (\pi n u)$ and correlation functions are
\begin{align*}
\mathbbm{E} [y_{n_1} \ldots y_{n_r}] = \sum_{\sigma} \int^1_0 d
u \prod_{j = 1}^r \frac{1}{2} \exp (i \pi \sigma_j 2^{n_j} u) = \frac{1}{2^r}
\sum_{\sigma} \delta (\sigma_1 2^{n_1} + \ldots + \sigma_r 2^{n_r})
\end{align*}
where the
sum is over the set $\{ (\sigma_1, \ldots, \sigma_r) | \sigma_i \in \{ - 1, 1
\} \}$ {\cite{beck_higher_1991}}. The only third order correlation function
that is non-zero is therefore $\mathbbm{E} [y^2_1 y_2] = \frac{1}{4}$.
This shows that, by Eq. \eqref{eq.c31_langevin},  $c^{(3,1)} = \frac{1}{4}$.

\begin{figure}[h]
	\centering
  \resizebox{225pt}{!}{\includegraphics{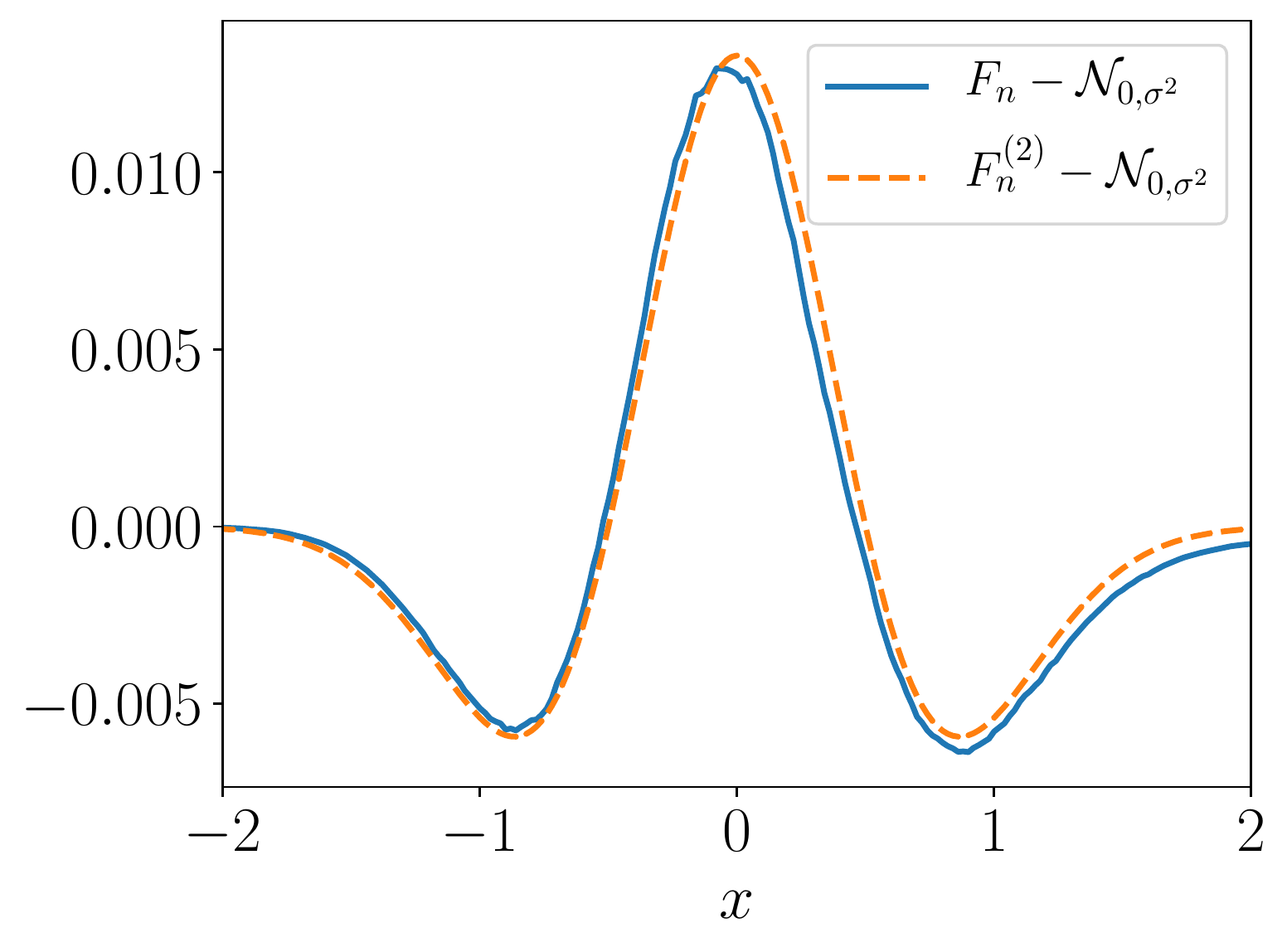}}
  \caption{The difference of the cumulative distribution function (cdf) $F_n(x) := \mathbb{P}\{x_n \leq x\}$ to the limiting
  \ Gaussian cumulative distribution function $\mathcal{N}_{0,\sigma^2}$ (solid line) and the
  difference of the cdf of the Edgeworth expansion $F_n^{(2)}(x) := \int_{-\infty}^{x} \rho_n^{(2)}(x) \, \mathrm{d} x$ to $\mathcal{N}_{0,\sigma^2}$
  (dashed line). Here $\tau = 0.01$, $n = 10^5$ and $10^{10}$ samples are 
  generated to estimate $F_n$.\label{fig.cheb_hist}}
\end{figure}

Figure \ref{fig.cheb_hist} shows that the first Edgeworth approximation
closely matches the deviations from Gaussianity observed in the distribution
of $x_n$ for large $n$ and small $\tau$.

\section{Conclusions}

In this paper we consider two applications of Edgeworth expansions.

Firstly, we have derived the Edgeworth coefficients of sums of dependent
random variables. To the author's knowledge, this is the first explicit derivation of this expansion in the literature. Equations for the expansion coefficients can be found in \cite{GoetzeHipp83}, however without derivation. Furthermore, the coefficient $c^{(4,1)}$ derived here differs substantially from the one found there. The numerical experiments in this manuscript corroborate the correctness of the expressions derived here. Furthermore,  they show the high accuracy of the Edgeworth approximation. This in turn supports the hypothesis that an Edgeworth expansion holds for this dynamical system, an assumption we have not proved here.

Secondly, we show that recent results on approximations of invariant distributions of slow-fast discrete maps fit into the general framework of Edgeworth
expansions. Approximations for the invariant distribution of the specific class of slow-fast linear Langeving maps have been derived in \cite{williams_stochastic_2018,beck_dynamical_1996} by different methods. The derivation given here puts these result in the context of the well-established topic of Edgeworth expansions. This provides a new view on these results and opens the way to extension to other classes of dynamical systems.

\section*{Acknowledgements}

The author would like to thank Georg Gottwald for stimulating and enjoyable
discussions.

\bibliographystyle{plain}\bibliography{Edgeworth.bib}

\appendix\section{Derivation of the Edgeworth expansion of
sums}\label{app.sums}

The aim is to derive expansions in orders of $\frac{1}{\sqrt{n}}$ of the cumulants of $x_n = \frac{1}{\sqrt{n}} \sum_{i = 1}^n f_1(y_i)$ as in Eq. (\ref{eq.cum-exp}). The expansion is most straightforwardly calculated after taking the $z$-transform w.r.t. $n$.


Taking the z-transform of the second moment $m^{(2)}_n \assign \mathbbm{E}
[x^2_n |x_0 = 0, y_0 \sim \rho_{\infty}]$
\begin{eqnarray*}
  \hat{m}^{(2)} (\lambda) : = \sum_{n = 0}^{\infty} \lambda^n m^{(2)}_n & = &
  \int \delta_{x_0} \otimes \rho_{\infty} ({\mathrm d} x, {\mathrm d} y)
  \sum_{n = 0}^{\infty} \lambda^n P^n x^2\\
  & = & \int \delta_{x_0} \otimes \rho_{\infty} ({\mathrm d} x, {\mathrm
  d} y) \frac{1}{1 - \lambda P} x^2
\end{eqnarray*}
where $P$ is the Koopman operator $P A (x, y) = A (x + \varepsilon f_1 (g
(y)), g (y))$ of the system
\begin{align*}
	x_{n+1} &= x_n + \varepsilon f_1(g(y_n)) \\
	y_n+1 &= g(y_n) \, .
\end{align*}
Note that in this system, setting $x_0=0$, we have $x_n = \sum_{k = 1}^n \varepsilon f_1(g(y))$. We will later be setting $\varepsilon = \frac{1}{\sqrt{n}}$ to obtain sums of the CLT form. The operator $P$ can be expanded as $P = \left( \sum_{k =
0}^{\infty} \frac{\varepsilon^k}{k!} P^k_x \right) P_y$ with $P_x A (x, y) =
f_1 (g (y)) \partial_x A (x, y)$ and $P_y A (x, y) = A (x, g (y))$.

Then since $(C - D)^{- 1} = C^{- 1} + C^{- 1} D C^{- 1} + C^{- 1} D C^{- 1} D
C^{- 1} + \ldots$ we have taking $C = 1 - \lambda P_y$ and $D = \lambda \left(
\sum_{k = 1}^{\infty} \frac{\varepsilon^k}{k!} P_x^k \right) P_y$
\begin{eqnarray*}
  \hat{m}^{(2)} (\lambda) & = & \int \delta_{x_0} \otimes \rho_{\infty}
  ({\mathrm d} x, {\mathrm d} y) \left( \frac{\lambda}{1 - \lambda} \left(
  \sum_{k = 1}^{\infty} \frac{\varepsilon^k}{k!} P_x^k \right) \frac{1}{1 -
  \lambda} \right. \\
  && \left. + \frac{\lambda}{1 - \lambda} \left( \sum_{k = 1}^{\infty}
  \frac{\varepsilon^k}{k!} P_x^k \right) \frac{\lambda P_y}{1 - \lambda P_y}
  \left( \sum_{k = 1}^{\infty} \frac{\varepsilon^k}{k!} P_x^k \right)
  \frac{1}{1 - \lambda} \right) x^2
\end{eqnarray*}
other terms in the expansion are zero since they have either not enough or too
many derivatives $\partial_x$. By the same reasoning, we can see that
\begin{eqnarray*}
  \hat{m}^{(2)} (\lambda) & = & \int \rho_{\infty} ({\mathrm d} y) \left(
  \frac{\lambda}{1 - \lambda} \left( \frac{\varepsilon^2}{2!} f_1^{} (g (y))^2
  \right) \frac{1}{1 - \lambda} \right. \\
  && \left. + \frac{\lambda}{1 - \lambda} (\varepsilon f_1^{} (g (y)))
  \frac{\lambda P_y}{1 - \lambda P_y} (\varepsilon f_1^{} (g (y))) \frac{1}{1
  - \lambda} \right) 2\\
  & = & \varepsilon^2 \frac{\lambda}{(1 - \lambda)^2} \mathbbm{E} [f_1^{}
  (y)^2] + 2 \varepsilon^2 \frac{\lambda}{(1 - \lambda)^2} \mathbbm{E}
  \left[ f_1^{} (y) \frac{\lambda P_y}{1 - \lambda P_y} f_1 (y) \right]
\end{eqnarray*}
where $\mathbbm{E} [A] = \int \rho_{\infty} ({\mathrm d} y) A (y)$ with $\rho_{\infty}$ the physical invariant measure of $y_{i+1} = g(y_i)$. We now
expand the Koopman operator as $P_y = p_0 + \sum_{i = 1}^{\infty} \alpha_i p_i$ , with $p_0 = | 1 \rangle \langle
\rho_{\infty} |$ and $p_i = | l_i \rangle \langle r_i |$,
where the left eigenfunctions ($1$ and $l_i$) and right eigenfunctions
($\rho_{\infty}$ and $r_i$) are mutually orthogonal. We then obtain
\begin{eqnarray*}
  \hat{m}^{(2)} (\lambda) & = & \varepsilon^2 \frac{\lambda}{(1 - \lambda)^2}
  \mathbbm{E} [f_1^{} (y)^2] + 2 \varepsilon^2 \frac{\lambda}{(1 -
  \lambda)^2} \sum_{i = 1}^{\infty} \frac{\lambda \alpha_i}{1 - \lambda
  \alpha_i} \mathbbm{E} [f_1^{} (y) | l_i \rangle \langle r_i | f_1 (y)]
\end{eqnarray*}

By the inverse z-transform (calculating the residue at $\lambda=1$ of $\hat{m}^{(2)}(\lambda)$) we have
\begin{eqnarray*}
  m^{(2)}_n & = & \varepsilon^2 n^{} \mathbbm{E} [f_1^{} (y)^2] + 2
  \varepsilon^2 \sum_{i = 1}^{\infty} \left( n \frac{\alpha_i}{1 - \alpha_i} -
  \frac{\alpha_i}{(1 - \alpha_i)^2} \right) \mathbbm{E} [f_1^{} (y) | l_i
  \rangle \langle r_i | f_1 (y)] \\
   & = & \varepsilon^2 n^{} \mathbbm{E} [f_1^{} (y)^2] + 2
   \varepsilon^2 n \mathbbm{E} \left[f_1^{} (y) \frac{P_y}{1-P_y} f_1 (y)] - 2
   \varepsilon^2
   \mathbbm{E} [f_1^{} (y) \frac{P_y}{(1-P_y)^2}  f_1 (y) \right]
\end{eqnarray*}
Noting that $\frac{P_y}{1-P_y} = \sum_{k = 1}^\infty P_y^k$ and $\frac{P_y}{(1-P_y)^2} = \sum_{k = 1}^\infty k P_y^k$, by setting
$\varepsilon = 1 / \sqrt{n}$, we have
\begin{eqnarray*}
   c^{(2)}_n = m^{(2)}_n & = & \left( \mathbbm{E} [f_1 (y)^2] + 2 \sum_{k = 1}^{\infty}
  \mathbbm{E} [f_1 (y) f_1 (g^k (y))] \right) + \frac{1}{n} \left( - 2
  \sum_{k = 1}^{\infty} k\mathbbm{E} [f_1 (y) f_1 (g^k (y))] \right)
  +\mathcal{O} \left( \frac{1}{n^2} \right)\\
  & = & \sigma^2 + \frac{1}{n} c^{(2,1)} +\mathcal{O} \left( \frac{1}{n^2} \right)
\end{eqnarray*}
with $\sigma^2$ as given in Eq. (\ref{eq.GK}) and $c^{(2,1)}$ as given in Eq. (\ref{eq.cum2}).

Similarly, we obtain for the third moment of $x_n$
\begin{eqnarray*}
 c^{(3)}_n = m^{(3)}_n &
  = & \frac{1}{\sqrt{n}} \left( \mathbbm{E} [f_1 (y)^3] + \sum_{k =
  1}^{\infty} 3 \left( \mathbbm{E} [f_1 (y)^2 f_1 (g^k (y))] +\mathbbm{E}
  \left[ f_1 (y)^{} f_1 (g^k (y))^2 \right] \right) \right. \\
  && \hspace*{.3\textwidth}\left. + 6 \sum_{k = 1}^{\infty} \sum_{l = 1}^{\infty} \mathbbm{E} [f_1 (y) f_1
  (g^k (y)) f_1 (g^{k + l} (y))] \right) \\
  &=& \frac{1}{\sqrt{n}} c^{(3,1)}
\end{eqnarray*}
with $c^{(3,1)}$ as in Eq. (\ref{eq.cum3}).

For the fourth moment, we have
\begin{eqnarray*}
  \hat{m}^{(4)} (\lambda) & = & \sum_{n = 0}^{\infty} \lambda^n m^{(4)}_n\\
  & = & \mathbbm{E} \left[ \frac{\lambda}{1 - \lambda} \left( f_1^4
  \frac{\varepsilon^4}{4!} \right) \frac{1}{1 - \lambda} \right] 4!\\
  &  & +\mathbbm{E} \left[ \frac{\lambda}{1 - \lambda} \left( f_1
  \frac{\varepsilon}{1!} \right) \frac{\lambda P_y}{1 - \lambda P_y} \left(
  f^3_1 \frac{\varepsilon^3}{3!} \right) \frac{1}{1 - \lambda} \right] 4!\\
  &  & +\mathbbm{E} \left[ \frac{\lambda}{1 - \lambda} \left( f^3_1
  \frac{\varepsilon^3}{3!} \right) \frac{\lambda P_y}{1 - \lambda P_y} \left(
  f_1 \frac{\varepsilon}{1!} \right) \frac{1}{1 - \lambda} \right] 4!\\
  &  & +\mathbbm{E} \left[ \frac{\lambda}{1 - \lambda} \left( f^2_1
  \frac{\varepsilon^2}{2!} \right) \frac{\lambda P_y}{1 - \lambda P_y} \left(
  f^2_1 \frac{\varepsilon^2}{2!} \right) \frac{1}{1 - \lambda} \right] 4!\\
  &  & +\mathbbm{E} \left[ \frac{\lambda}{1 - \lambda} \left( f_1
  \frac{\varepsilon}{1!} \right) \frac{\lambda P_y}{1 - \lambda P_y} \left(
  f_1 \frac{\varepsilon}{1!} \right) \frac{\lambda P_y}{1 - \lambda P_y}
  \left( f^2_1 \frac{\varepsilon^2}{2!} \right) \frac{1}{1 - \lambda} \right]
  4!\\
  &  & +\mathbbm{E} \left[ \frac{\lambda}{1 - \lambda} \left( f_1
  \frac{\varepsilon}{1!} \right) \frac{\lambda P_y}{1 - \lambda P_y} \left(
  f^2_1 \frac{\varepsilon^2}{2!} \right) \frac{\lambda P_y}{1 - \lambda P_y}
  \left( f_1 \frac{\varepsilon}{1!} \right) \frac{1}{1 - \lambda} \right] 4!\\
  &  & +\mathbbm{E} \left[ \frac{\lambda}{1 - \lambda} \left( f^2_1
  \frac{\varepsilon^2}{2!} \right) \frac{\lambda P_y}{1 - \lambda P_y} \left(
  f_1 \frac{\varepsilon}{1!} \right) \frac{\lambda P_y}{1 - \lambda P_y}
  \left( f_1 \frac{\varepsilon}{1!} \right) \frac{1}{1 - \lambda} \right] 4!\\
  &  & +\mathbbm{E} \left[ \frac{\lambda}{1 - \lambda} \left( f_1
  \frac{\varepsilon}{1!} \right) \frac{\lambda P_y}{1 - \lambda P_y} \left(
  f_1 \frac{\varepsilon}{1!} \right) \frac{\lambda P_y}{1 - \lambda P_y}
  \left( f_1 \frac{\varepsilon}{1!} \right) \frac{\lambda P_y}{1 - \lambda
  P_y} \left( f_1 \frac{\varepsilon}{1!} \right) \frac{1}{1 - \lambda} \right]
  4!\\
  & = & \varepsilon^4 \frac{\lambda}{(1-\lambda)^2} \mathbbm{E} [f_1^4]\\
  &  & + 4 \varepsilon^4 \frac{\lambda}{(1 - \lambda)^2} \sum_{i = 1}^\infty \mathbbm{E} \left[
  f_1 \left( \frac{\lambda \alpha_i}{1 - \lambda \alpha_i} p_i \right) f^3_1
  \right]\\
  &  & + 4 \varepsilon^4 \frac{\lambda}{(1 - \lambda)^2} \sum_{i = 1}^\infty \mathbbm{E} \left[
  f^3_1 \left( \frac{\lambda \alpha_i}{1 - \lambda \alpha_i} p_i \right) f_1
  \right]\\
  &  & + 6 \varepsilon^4 \frac{\lambda}{(1 - \lambda)^2} \mathbbm{E} \left[
  f^2_1 \left( \frac{\lambda}{1 - \lambda} p_0 + \sum_{i = 1}^\infty \frac{\lambda \alpha_i}{1 -
  \lambda \alpha_i} p_i \right) f^2_1 \right]\\
  &  & + 12 \varepsilon^4 \frac{\lambda}{(1 - \lambda)^2} \mathbbm{E} \left[
  f_1 \left( \sum_{i = 1}^\infty \frac{\lambda \alpha_i}{1 - \lambda \alpha_i} p_i \right) f_1
  \left( \frac{\lambda}{1 - \lambda} p_0 + \sum_{j = 1}^\infty \frac{\lambda \alpha_j}{1 - \lambda
  \alpha_j} p_j \right) f^2_1 \right]\\
  &  & + 12 \varepsilon^4 \frac{\lambda}{(1 - \lambda)^2} \mathbbm{E} \left[
  f_1 \left( \sum_{i = 1}^\infty \frac{\lambda \alpha_i}{1 - \lambda \alpha_i} p_i \right) f^2_1
  \left( \sum_{j = 1}^\infty \frac{\lambda \alpha_j}{1 - \lambda \alpha_j} p_j \right) f_1
  \right]\\
  &  & + 12 \varepsilon^4 \frac{\lambda}{(1 - \lambda)^2} \mathbbm{E} \left[
  f^2_1 \left( \frac{\lambda}{1 - \lambda} p_0 + \sum_{i = 1}^\infty \frac{\lambda \alpha_i}{1 -
  \lambda \alpha_i} p_i \right) f_1 \left( \sum_{j = 1}^\infty \frac{\lambda \alpha_j}{1 - \lambda
  \alpha_j} p_j \right) f_1 \right]\\
  &  & + 24 \varepsilon^4 \frac{\lambda}{(1 - \lambda)^2} \mathbbm{E} \left[
  f_1 \left( \sum_{i = 1}^\infty \frac{\lambda \alpha_i}{1 - \lambda \alpha_i} p_i \right) f_1
  \left( \frac{\lambda}{1 - \lambda} p_0 + \sum_{j = 1}^\infty \frac{\lambda \alpha_j}{1 - \lambda
  \alpha_j} p_j \right) f_1 \left( \sum_{k = 1}^\infty \frac{\lambda \alpha_k}{1 - \lambda
  \alpha_k} p_k \right) f_1 \right]\\
\end{eqnarray*}
\begin{eqnarray*}
  & = & \varepsilon^4 \frac{\lambda}{(1 - \lambda)^2} \mathbbm{E} [f_1^4]\\
  &  & + 6 \varepsilon^4 \frac{\lambda^2}{(1 - \lambda)^3} \mathbbm{E}
  [f_1^2]^2\\
  &  & + 2 \varepsilon^4 \frac{\lambda}{(1 - \lambda)^2} \sum_{i=1}^\infty \frac{\lambda
  \alpha_i}{1 - \lambda \alpha_i} (2\mathbbm{E} [f_1 p_i f^3_1] + 2\mathbbm{E}
  [f^3_1 p_i f_1] + 3\mathbbm{E} [f^2_1 p_i f^2_1])\\
  &  & + 12 \varepsilon^4 \frac{\lambda^2}{(1 - \lambda)^3} \sum_{i=1}^\infty  \frac{\lambda
  \alpha_i}{1 - \lambda \alpha_i} (2\mathbbm{E} [f_1 p_i f_1] \mathbbm{E}
  [f_1^2])\\
  &  & + 12 \varepsilon^4 \frac{\lambda}{(1 - \lambda)^2} \sum_{i,j=1}^\infty  \frac{\lambda
  \alpha_i}{1 - \lambda \alpha_i}  \frac{\lambda \alpha_j}{1 - \lambda
  \alpha_j}  (\mathbbm{E} [f_1 p_i f_1 p_j f^2_1] +\mathbbm{E} [f_1 p_i f^2_1
  p_j f_1] +\mathbbm{E} [f^2_1 p_i f_1 p_j f_1])\\
  &  & + 24 \varepsilon^4 \frac{\lambda^2}{(1 - \lambda)^3} \sum_{i,j=1}^\infty  \frac{\lambda
  \alpha_i}{1 - \lambda \alpha_i}  \frac{\lambda \alpha_j}{1 - \lambda
  \alpha_j} \mathbbm{E} [f_1 p_i f_1] \mathbbm{E} [f_1 p_j f_1]\\
  &  & + 24 \varepsilon^4 \frac{\lambda}{(1 - \lambda)^2}  \sum_{i,j,k=1}^\infty \frac{\lambda
  \alpha_i}{1 - \lambda \alpha_i}  \frac{\lambda \alpha_j}{1 - \lambda
  \alpha_j} \frac{\lambda \alpha_k}{1 - \lambda \alpha_k} \mathbbm{E} [f_1 p_i
  f_1 p_j f_1 p_k f_1]
\end{eqnarray*}
By inverse z-transform of $\hat{m}^{(4)}(\lambda)$, calculating the residue of $\hat{m}^{(4)}(\lambda) \lambda^{-n-1}$ at $\lambda = 1$ we obtain $m^{(4)}_n$. Note that there are also poles at $1 /
\alpha_i$, but these contribute terms of order $\alpha^n_i$, which decay exponentially with $n$ and therefore don't appear in the Edgeworth expansion.
\begin{align*}
  m^{(4)}_n  = & \varepsilon^4 n\mathbbm{E} [f_1^4]\\
   & + 6 \varepsilon^4 \frac{1}{2} (n^2 - n) \mathbbm{E} [f_1^2]^2\\
  & + 2 \varepsilon^4 n \sum_{i = 1}^\infty \frac{\alpha_i}{1 - \alpha_i} (2\mathbbm{E} [f_1
  p_i f^3_1] + 2\mathbbm{E} [f^3_1 p_i f_1] + 3\mathbbm{E} [f^2_1 p_i
  f^2_1])\\
  & + 12 \varepsilon^4 \frac{1}{2} \sum_{i = 1}^\infty \left( n^2 \frac{\alpha_i}{1 -
  \alpha_i} - n \left( \frac{\alpha_i}{1 - \alpha_i} + 2 \frac{\alpha_i}{(1 -
  \alpha_i)^2} \right) \right) (2\mathbbm{E} [f_1 p_i f_1] \mathbbm{E}
  [f_1^2])\\
  & + 12 \varepsilon^4 n \sum_{i,j = 1}^\infty \frac{\alpha_i}{1 - \alpha_i} \frac{\alpha_j}{1 -
  \alpha_j} (\mathbbm{E} [f_1 p_i f_1 p_j f^2_1] +\mathbbm{E} [f_1 p_i f^2_1
  p_j f_1] +\mathbbm{E} [f^2_1 p_i f_1 p_j f_1])\\
  & + 24 \varepsilon^4 \frac{1}{2} \sum_{i,j = 1}^\infty \left( n^2 \frac{\alpha_i}{1 -
  \alpha_i}  \frac{\alpha_j}{1 - \alpha_j} \right. \\
  & \left. - n \left( \frac{\alpha_i}{1 - \alpha_i}  \frac{\alpha_j}{1
  - \alpha_j} + 2 \frac{\alpha_i}{(1 - \alpha_i)^2} \frac{\alpha_j}{1 -
  \alpha_j} + 2 \frac{\alpha_i}{1 - \alpha_i}  \frac{\alpha_j}{(1 -
  \alpha_j)^2} \right) \right) \mathbbm{E} [f_1 p_i f_1] \mathbbm{E} [f_1 p_j
  f_1]\\
  & + 24 \varepsilon^4 n \sum_{i,j,k = 1}^\infty \frac{\alpha_i}{1 - \alpha_i}  \frac{\alpha_j}{1 -
  \alpha_j} \frac{\alpha_k}{1 - \alpha_k} \mathbbm{E} [f_1 p_i f_1 p_j f_1 p_k
  f_1]
\end{align*}
\begin{align*}
  = & \varepsilon^4 n^2 \left( 3\mathbbm{E} [f_1^2]^2 + 12\mathbbm{E}
  \left[ f_1 \frac{P_y}{1 - P_y} f_1 \right] \mathbbm{E} [f_1^2] + 12\mathbbm{E}
  \left[ f_1 \frac{P_y}{1 - P_y} f_1 \right]^2 \right)\\
  &  + \varepsilon^4 n \left( \mathbbm{E} [f_1^4] - 3\mathbbm{E}
  [f_1^2]^2 + 4\mathbbm{E} \left[ f_1 \frac{P_y}{1 - P_y} f^3_1 \right] +
  4\mathbbm{E} \left[ f^3_1 \frac{P_y}{1 - P_y} f_1 \right] + 6\mathbbm{E} \left[
  f^2_1 \frac{P_{\perp}}{1 - P_{\perp}} f^2_1 \right] \right. \\
  &- 12\mathbbm{E} \left[ f_1 \frac{P_y}{1 - P_y} f_1 \right] \mathbbm{E} [f_1^2]
  - 24\mathbbm{E} \left[ f_1 \frac{P_y}{(1 - P_y)^2} f_1 \right] \mathbbm{E}
  [f_1^2] \\
  &+ 12 \left( \mathbbm{E} \left[ f_1 \frac{P_y}{1 - P_y} f_1 \frac{P_{\perp}}{1 -
  P_{\perp}} f^2_1 \right] +\mathbbm{E} \left[ f_1 \frac{P_y}{1 - P_y} f^2_1
  \frac{P_y}{1 - P_y} f_1 \right] +\mathbbm{E} \left[ f^2_1 \frac{P_{\perp}}{1 -
  P_{\perp}} f_1 \frac{P_y}{1 - P_y} f_1 \right] \right) \\
  &- 12 \left( \mathbbm{E} \left[ f_1 \frac{P_y}{1 - P_y} f_1 \right] \mathbbm{E}
  \left[ f_1 \frac{P_y}{1 - P_y} f_1 \right] + 4\mathbbm{E} \left[ f_1 \frac{P_y}{(1
  - P_y)^2} f_1 \right] \mathbbm{E} \left[ f_1 \frac{P_y}{1 - P_y} f_1 \right]
  \right) \\
  &\left. + 24\mathbbm{E} \left[ f_1 \frac{P_y}{1 - P_y} f_1 \frac{P_{\perp}}{1 -
  P_{\perp}} f_1 \frac{P_y}{1 - P_y} f_1 \right] \right)\\
  = & 3 \varepsilon^4 n^2 \sigma^4\\
  & + \varepsilon^4 n \left( \mathbbm{E} [f_1^4] + 4\mathbbm{E} \left[
  f_1 \frac{P_y}{1 - P_y} f^3_1 \right] + 4\mathbbm{E} \left[ f^3_1 \frac{P_y}{1 -
  P_y} f_1 \right] + 6\mathbbm{E} \left[ f^2_1 \frac{P_{\perp}}{1 - P_{\perp}}
  f^2_1 \right] \right. \\
  & + 12 \left( \mathbbm{E} \left[ f_1 \frac{P_y}{1 - P_y} f_1 \frac{P_{\perp}}{1 -
  P_{\perp}} f^2_1 \right] +\mathbbm{E} \left[ f_1 \frac{P_y}{1 - P_y} f^2_1
  \frac{P_y}{1 - P_y} f_1 \right] +\mathbbm{E} \left[ f^2_1 \frac{P_{\perp}}{1 -
  P_{\perp}} f_1 \frac{P_y}{1 - P_y} f_1 \right] \right) \\
  & \left. + 24\mathbbm{E} \left[ f_1 \frac{P_y}{1 - P_y} f_1 \frac{P_{\perp}}{1 -
  P_{\perp}} f_1 \frac{P_y}{1 - P_y} f_1 \right] - 3 \sigma^4 + 12 \sigma^2
  c^{(2,1)} \right)
\end{align*}
where $P_\perp = \sum_{i = 1}^\infty p_i $.

Finally, setting $\varepsilon = \frac{1}{\sqrt{n}}$ and noting the $P^k_\perp=P^k - p_0$, we get for the fourth cumulant $c^{(4)}_n = m^{(4)}_n - 3 (m^{(2)}_n)^2 = m^{(4)}_n - 3 (\sigma^2 + \frac{c^{(2,1)}}{n})^2 = \frac{1}{n} c^{(4,1)}$ with $c^{(4,1)}$ as given in Eq. (\ref{eq.cum4}).

\section{Convergence of cumulants for the tripling map}

Here we present additional evidence of the validity of the cumulant expansion of Eq. (\ref{eq.cum-exp}) for sums  $x_n = \frac{1}{\sqrt{n}} \sum_{i = 1}^n f_1(y)$ with $f_1(y) = y^5 + y^4 - \frac{1}{6}  - \frac{1}{5}$ of the tripling map $y_{i+1} = 3 y_i \mod 1$.

\begin{figure}[h]
	\centering
	\begin{tabular}{ c c}
		\resizebox{175pt}{!}{ \includegraphics{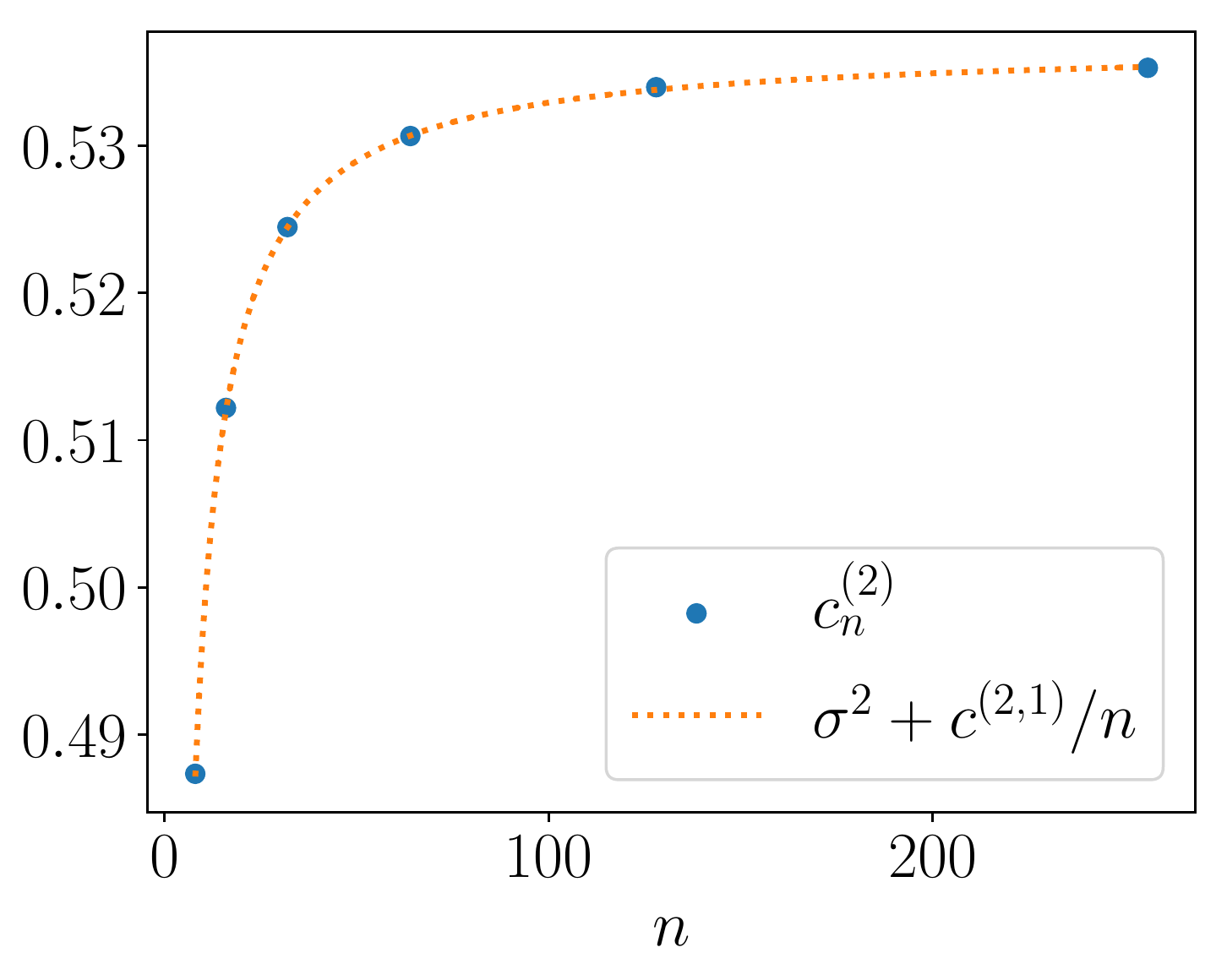}} & \resizebox{175pt}{!}{\includegraphics{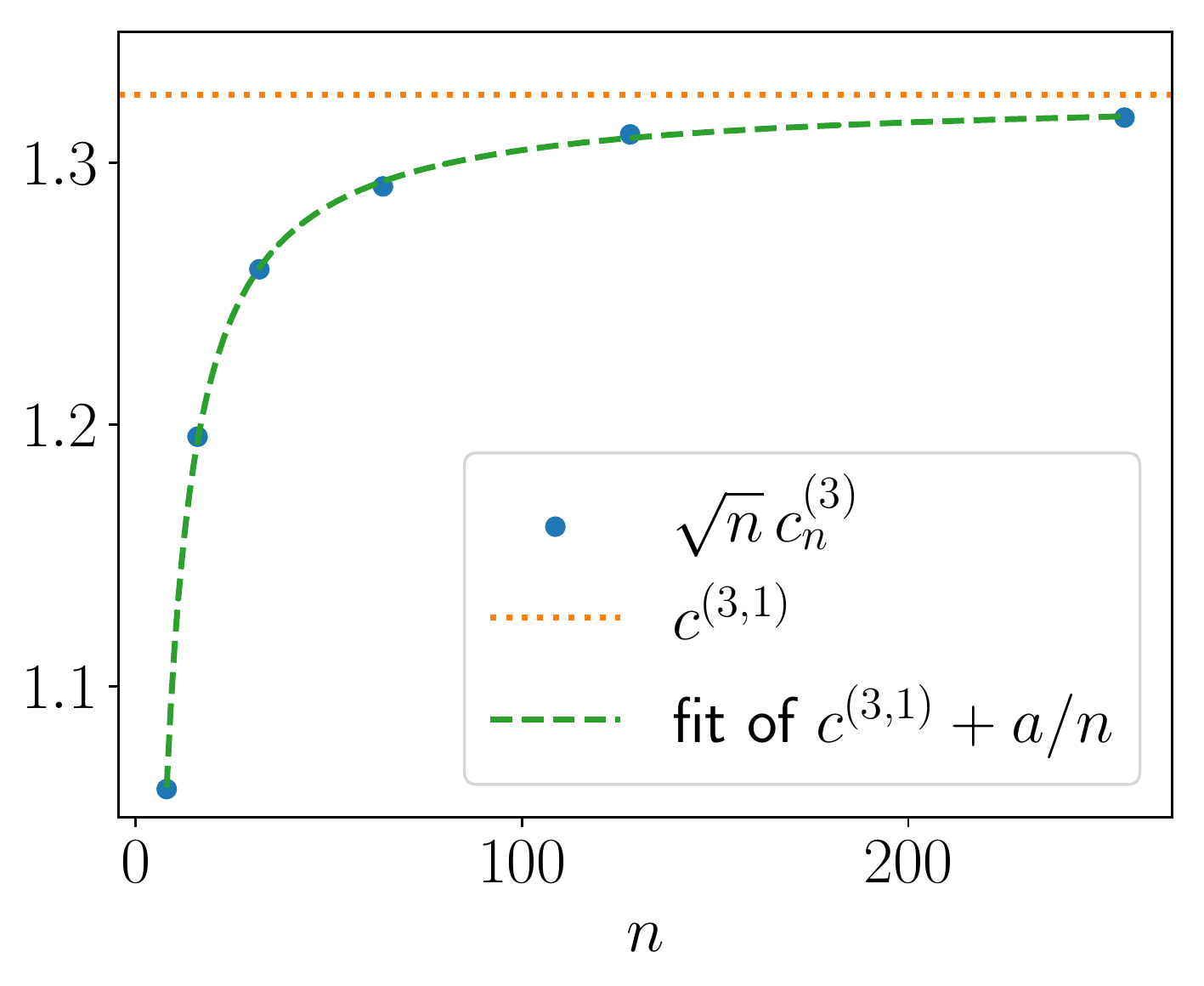}} \\
		(a) & (b) \\
		\resizebox{175pt}{!}{\includegraphics{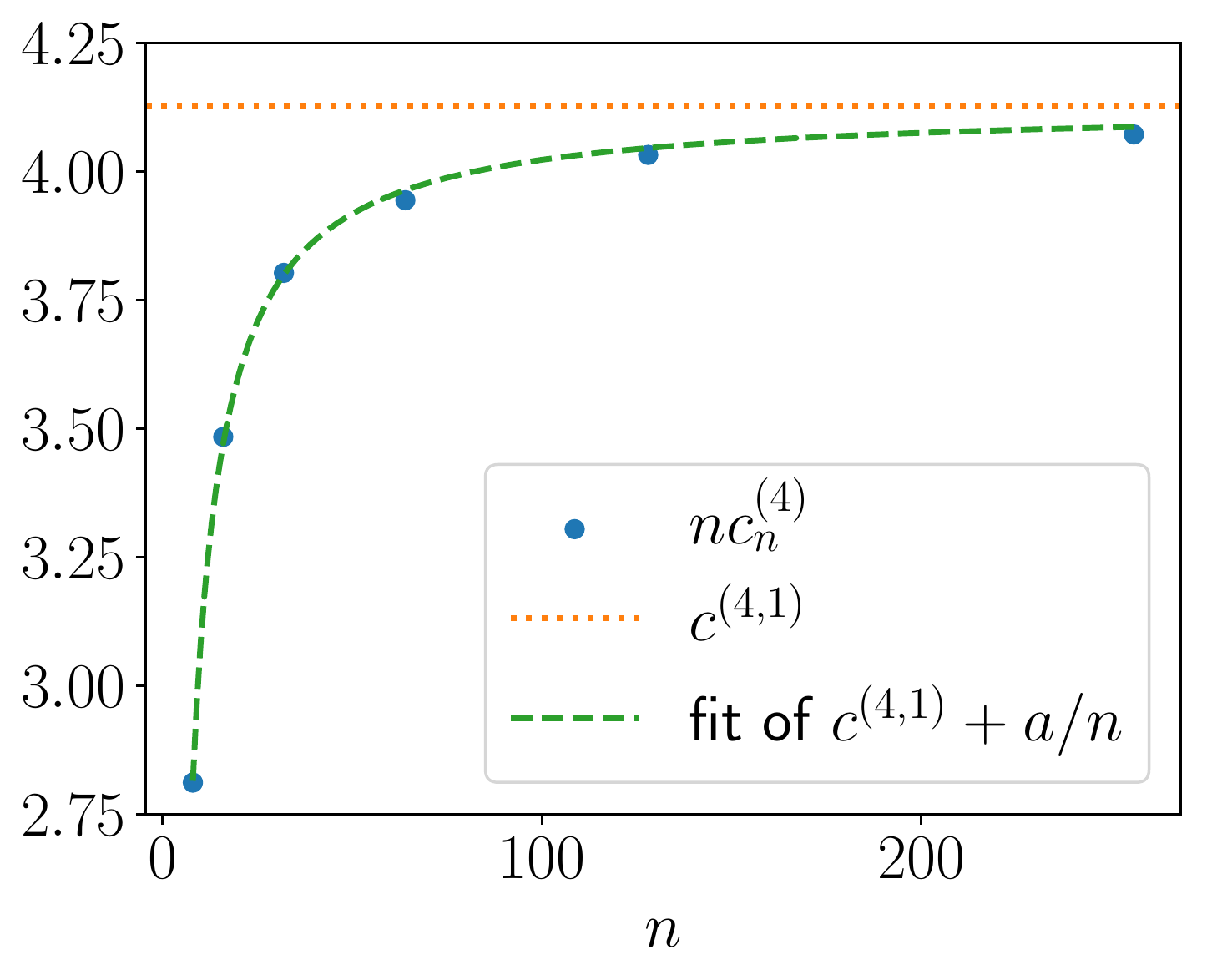}} & \\
		(c) &
	\end{tabular}
	 
  \caption{Cumulants of sums of the tripling map, comparing Monte Carlo simulation (blue dots) with analytic asymptotics (yellow dotted lines) and a fit of the analytic asymptotics with one higher order term (green dashed lines) for (a) the second cumulant, (b) the third cumulant and (c) the fourth cumulant.}
  \label{fig.tripling_cums}
\end{figure}

Figure \ref{fig.tripling_cums} demonstrates that the cumulants of $x_n$ indeed vary with $n$ as described in Eqs. (\ref{eq.cum-exp}). The values for $\sigma^2$, $c^{(2,1)}$, $c^{(3,1)}$ and $c^{(4,1)}$ analytically derived here (see Eqs. (\ref{eq.cum2})-(\ref{eq.cum4})) indeed give the leading order asymptotics of these cumulants. Furthermore, we show for the third and fourth cumulants that by including a higher order correction of $1/n$, the numerical values are matched extremely well. An analytic expression for this higher order correction is not derived here, but could be found be the same techniques developed here.

\section{SageMath code to calculate tripling map cumulant expansion}
\label{app.sage}
Tested in SageMath version 8.6, release date 2019-01-15.

\begin{lstlisting}
k, n, p = var('k,n,p')
assume(p>1)
assume(n,'real')
assume(n>0)

# observable
A(x) = x^5 + x^4
A(x) = A(x) - integral(A,x,0,1)

# second order correlation
C(n,p) = sum(integral((A(x))*(A((p^n)*x - k)), x, k/(p^n),
				(k+1)/(p^n)),k,0,(p^n) - 1)
C(n) = C(n,3)

# Green-Kubo equation
sigma2 = C(0) + 2*sum( (C(n)).expand().simplify(),
			n, 1, oo, algorithm='giac')
# First correction to second cumulant
c21 = - 2*sum((n*C(n)).expand(),n,1,oo, algorithm='giac')

# third order correlations
k1, k2, n1, n2 = var('k1,k2,n1,n2', domain="positive")
C3(n1,n2) = sum(sum(integral((A(x))*(A((3^n1)*x - (k1-1)))
    *(A((3^(n1+n2))*x - (3^n2)*(k1-1) - (k2-1))), 
    x, (k1-1)/(3^n1) + (k2-1)/(3^(n1+n2)), (k1-1)/(3^n1) + k2/(3^(n1+n2))),
                    k1, 1, 3^n1), k2, 1, 3^n2)

# first correction to the third cumulant
c31 = (C3(0,0) + 3*sum(C3(n1,0).expand(),n1,1,oo, algorithm='giac')
    + 3*sum(C3(0,n2).expand(),n2,1,oo, algorithm='giac')
    + 6*sum(sum(C3(n1,n2).expand(),n2,1,oo, algorithm='giac'),
            n1,1,oo, algorithm='giac'))

# fourth order correlations
k1, k2, k3, n1, n2, n3 = var('k1,k2,k3,n1,n2,n3', domain="positive")
C4(n1,n2,n3) = sum(sum(sum(integral((A(x))*(A((3^n1)*x - (k1-1)))
    *(A((3^(n1+n2))*x - (3^n2)*(k1-1) - (k2-1)))
    *(A((3^(n1+n2+n3))*x - (3^(n2+n3))*(k1-1) - 3^n3*(k2-1) - (k3-1))),
    x, (k1-1)/(3^n1) + (k2-1)/(3^(n1+n2)) + (k3-1)/(3^(n1+n2+n3)),
    (k1-1)/(3^n1) + (k2-1)/(3^(n1+n2)) + k3/(3^(n1+n2+n3))), k1, 1, 3^n1),
    k2, 1, 3^n2), k3, 1, 3^n3)

# first correction to the fourth cumulant
algo = 'sympy'
c41a = C4(n1=0,n2=0,n3=0)
c41b = sum(C4(n1=n1,n2=0,n3=0).expand(),n1,1,oo,algorithm=algo)
c41c = sum(C4(n1=0,n2=0,n3=n3).expand(),n3,1,oo,algorithm=algo)
c41d = sum((C4(n1=0,n2=n2,n3=0) - C(0)^ 2).expand(),n2,1,oo,algorithm=algo)
c41e = sum(sum((C4(n1=n1,n2=n2,n3=0) - C(n1)*C(0)).expand(),
          n1,1,oo,algorithm=algo),n2,1,oo,algorithm=algo)
c41f = sum(sum((C4(n1=n1,n2=0,n3=n3)).expand(),
          n1,1,oo,algorithm=algo),n3,1,oo,algorithm=algo)
c41g = sum(sum((C4(n1=0,n2=n2,n3=n3) - C(0)*C(n3)).expand(),
          n3,1,oo,algorithm=algo),n2,1,oo,algorithm=algo)
c41h = sum(sum(sum((C4(n1=n1,n2=n2,n3=n3) - C(n1)*C(n3)).expand(),
          n2,1,oo,algorithm=algo).expand(),n1,1,oo,algorithm=algo).expand(),
          n3,1,oo,algorithm=algo)

c41 = c41a + 4*c41b + 4*c41c + 6*c41d + 12*(c41e + c41f + c41g) + 24*c41h
          - 3*sigma2^2 + 6*sigma2*c21
\end{lstlisting}


\end{document}